\documentclass[twocolumn,prb,showpacs,superscriptaddress,superscriptaddress,floats]{revtex4}
\usepackage{calc}
\usepackage{graphicx}
\begin{document}
\def \beq{\begin{equation}}
\def \eeq{\end{equation}}

\begin{abstract}
We argue that topological meron excitations, which are in a strong
coupling phase (bound in pairs) in infinite quantum Hall ferromagnets, become
deconfined in finite size quantum Hall systems. Although
effectively for larger systems meron energy grows with the size of
the system, when gyromagnetic ratio is small
meron  becomes the lowest lying  state
of a quantum Hall droplet. This comes as a consequence of the
many-body correlations built in the meron construction that minimize
the interaction energy. We demonstrate this by using  mean field
ansatzes for meron wave function. The ansatzes will enable us to
consider much larger system sizes than in the previous work [A.
Petkovi\'{c} and M.V. Milovanovi\'{c}, PRL {\bf 98}, 066808 (2007)],
where fractionalization into merons was introduced.
\end{abstract}


\title{Meron ground states of quantum Hall droplets}
\author{M.V. Milovanovi\'{c}}
\affiliation{Institute of Physics, P.O.Box 68, 11080 Belgrade,
Serbia}
\author{E. Dobard\v{z}i\'{c}} \affiliation{Department of Physics, University of
Belgrade, P.O. Box 368, 11001 Belgrade, Serbia}
\author{Z. Radovi\'{c}} \affiliation{Department of Physics, University of
Belgrade, P.O. Box 368, 11001 Belgrade, Serbia}

\date{\today}
\maketitle \vskip2pc
\narrowtext
\section{Introduction}
A quantum dot (QD) in high magnetic fields,~\cite{dot} so-called
quantum Hall droplet,~\cite{drop} represents an exciting playground
for correlation effects in interacting electron systems. Some
possible effects can be found in transport measurements that detect
oscillations in magnetoconductance \cite{pmc}
 in the interval  $ 2 \geq \nu \geq 1 $ of
filling factors.  The associated minima of the current
amplitude are most completely understood taking
into account a tendency of the system to find itself in
 {\em depolarized}, highly correlated ground states at some fractions in between despite Zeeman
cost.\cite{haw,wh,im,kh}

In this paper we will introduce closely related  depolarized  states
as meron topological excitations of the $\nu = 1$ completely
polarized droplet, therefore extending topological models to a
finite system. They will represent relevant lowest lying - ground
state configurations of the droplet when gyromagnetic ratio is
small.

Whenever we think about quasiparticles in small systems we are
skeptical about their existence or clear cut description that we can
find in infinite systems. That is even more true if we deal with
quasiparticles which quantization can be based on topological
considerations like skyrmions in quantum Hall (QH) ferromagnets.
\cite{so} Nevertheless, as we will argue here, topological objects
like merons, a meron is a half of skyrmion,~\cite{moo} can exist as
lowest lying states of a  quantum Hall droplet. Therefore
fractionalization and quantization, characteristic to QH systems,
may persist even in small spin unpolarized systems bringing
topological objects to their description.

The spin characterization of QD states at realistic Zeeman coupling
is an open problem.~\cite{sil} A very thorough understanding of the
completely polarized case exists that includes a description of the
QD states by the way of a vortex
quantization.~\cite{sat,tap,tetal,betal} The introduction of higher
angular momentum states or the increase of magnetic field with
respect to the most compact, maximum density configuration of a dot
is followed by vortex appearance inside the dot. On the other hand
there is a need to introduce a classification to various (partially
polarized) states at arbitrary or realistic Zeeman
coupling.~\cite{zee}

In the previous work, Ref.~\onlinecite{pm}, by one of the present
authors, the Coulomb interaction problem of small quantum Hall
droplets (dots) with $N = 4$ and $N = 6$, $N$ is the number of
electrons, in the limit of zero Zeeman coupling, was studied by
exact diagonalization, in the lowest Landau level approximation. It
was shown that the lowest lying states of these small quantum Hall
droplets can be described and classified as states of merons.

In this work we address the question of the existence of meron
ground states in large quantum Hall droplets, $N \sim 20$, that
exact diagonalizations can not reach. At the same time, by
extrapolation, we will be able to estimate the size of the quantum
Hall droplet, $N \sim 100$, at which meron confinement takes place
i.e. when we can not expect meron ground states. We will be able to
do this by using a model wave function that describes a meron of
arbitrary winding number positioned at the center of a droplet. In
this sense our approach is variational, takes into account a mean
field description of a meron, and compares it energetically first
with the ground state - maximum density droplet (MDD) spin polarized
configuration, but also with other meron configurations - of
different winding number.

\section{DESCRIPTION OF MERON GROUND STATES}
At the heart of the theory of the quantum Hall systems at effective
small gyromagnetic ratio, and filling factor $\nu = 1$, i.e. quantum
Hall ferromagnets, is the commensuration of spin and charge
deviations from the ground state values.~\cite{so} Exchange
interaction prefers smooth tumbling of spins that follows changes in
the charge distribution. Topology plays an important role in the
theory of infinite size quantum Hall ferromagnets.~\cite{so} After
the spontaneous symmetry breaking the polarization of the ground
state is in a definite direction and fixes the boundary condition at
infinite radius. Due to then possible mapping between real and
internal (spin) spaces, the skyrmion excitations with nontrivial
value of topological charge, and the same value of electric charge
implied by the commensuration, can be identified. This classical
(nonlinear-$\sigma$ model) skyrmion solution finds a concrete
quantum-mechanical realization in the following construction
\cite{moo}
\begin{equation}
\label{sky} \Psi_{s}  = \prod_{i=1}^{N} \left[ \begin{array}{c}
\lambda \\ z_{i} \end{array} \right] \prod_{i<j} (z_{i} - z_{j}).
\end{equation}
We omitted Gaussian factors and $z$'s are 2D coordinates of $N$
electrons in the lowest Landau level (LLL) in units of the magnetic length,
$l_B=\sqrt{\hbar c/eB}$. In this construction the
coordinates (in the orbital part) participate in the spinor part.
The complex number $\lambda$ denotes the size of the skyrmion
signifying the characteristic length when the spin, which is around
the center pointing up, starts gradually pointing down as it does at
infinity ( $|z| \rightarrow \infty $). The parameter $\lambda$ is
finite , determined by the ratio between the strength of the
interaction and the Zeeman coupling.

Now we specialize to the case of a droplet. We simplify the matters
assuming that the gyromagnetic factor  is zero. Thus we are in a
scale invariant situation where $\lambda$ may be function only of
the size ($N$) of the system. We take $\lambda \rightarrow c \equiv
\sqrt{N}$,~\cite{mvm} and consider constructions for merons of the
following form
\begin{equation}
\label{mnn}
\Psi_{m}  = \prod_{i=1}^{N}
\left[ \begin{array}{c}  c^{n} \\ z_{i}^{n} \end{array} \right]
\prod_{i<j} (z_{i} - z_{j}),
\end{equation}
where $n$ is a positive integer. The mean angular momentum $M$ can
be easily (numerically) calculated and with respect to the ground
state value, $ M_{o} = N (N - 1)/2$, we find that
\begin{equation}
\Delta M = M - M_{o} = n \frac{N}{2} + \frac{n (n - 1)}{2}
\label{mom}
\end{equation}
 approximate very well
the calculated $\Delta M$, the better, the
larger $N$ is (Fig. 1).
\begin{figure}
\includegraphics[width= 7cm]{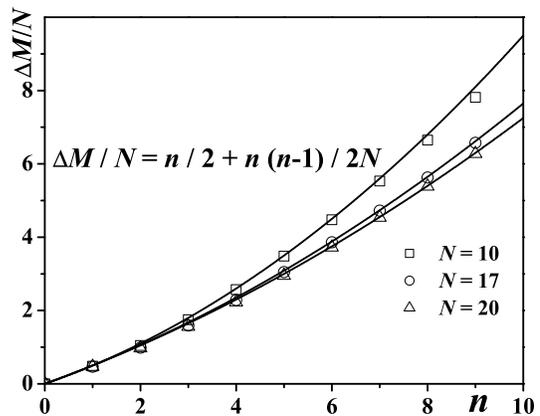}
\caption{The mean angular momentum with respect to the ground state value of
 the constructions in Eq.(\ref{mnn}). The results for various system sizes,
 i.e. number of electrons ($N$), are very well fitted with Eq.(\ref{mom}).
 The fit is better the larger the system size is.}
\end{figure}
Therefore, for $n = 1$, the excitation, in the mean, is the
excitation of one half of the flux quantum, $\Delta M = N $ being
the excitation of (Laughlin's  quasi) hole of one flux quantum or a
vortex. A skyrmion is a generalization of a Laughlin quasihole in
the case of quantum Hall ferromagnets. We may then identify the
construction with a classical one for a single meron being "half of
the skyrmion", and carrying flux of one half of one flux
quantum ($h c/e$). Also its magnetization, from pointing up in the
center, gradually transforms into one in the plane ($S_z=0$) on the
edges resembling the half of the skyrmion.~\cite{moo} In the case of
a skyrmion the magnetization would proceed to transform into configuration that points down on the
edges with equal magnitude but opposite direction with respect
to the configuration at the center. For a detailed description of
the single meron (Eq. (\ref{mnn}) with $n=1$) see below and Fig. 5.

Now we will show the results of the calculations of the total (interaction
plus confining) energy with respect to a suitable Hamiltonian of the states expressed
by Eq.(\ref{mnn}).

The interaction part of the Hamiltonian we work with can be described as a truncated
pseudopotential interaction \cite{hal} with only pseudopotentials $V_{l}$ for $l = 0, 1 $
nonzero and positive. Here $l$ is the relative angular momentum of a pair of
electrons in the LLL. Precisely,
\begin{equation}
H_{I} = \sum_{i < j} \sum_{l = 0}^{1} V_{l} P_{ij}^{l},
\label{iham}
\end{equation}
where $ P_{ij}^{l}$ is the projector of a state in the LLL of the
pair $(ij)$ to the definite relative angular momentum value $l$. We
consider $V_{0}$ much larger than $V_{1}$, making the state in
Eq.(\ref{mnn}) with $n = 0$ a unique ground state (up to SU(2)
rotation) of the system at filling factor one. In addition to $V_{0}
\neq 0$ we choose $V_{1} \neq 0$ because we want to break  energy
degeneracy among topological constructions implied by the hard core
(only $V_{0} \neq 0$) model.~\cite{mac} For an explanation how
to implement the model interaction with pseudopotentials
see Appendix. That meron excitations
(described in mean field as central constructions by Eq.(\ref{mnn}))
are indeed lowest lying states of a small droplet was shown in
Ref.~\onlinecite{pm} by exact diagonalization  in the case of the Coulomb
interaction.

The interaction energy of the ground state we calculated taking
$V_{1} = 1$ in $H_{I}$ and the results are plotted in Fig. 2.
\begin{figure}
\includegraphics[width= 7cm]{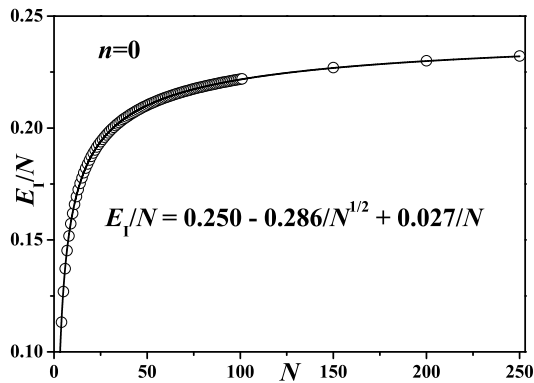}
\caption{The ground state interaction energy per electron in units
of $V_1$. The analytical fit is an expansion in inverse powers of $\sqrt{N}$
 which is the measure of the radius of the system.}
\end{figure}
We see the dependence of the interaction energy per particle on the
size - the number of particles in the system. The energy can be very
accurately fitted with the analytical expression in the Figure as an
expansion in inverse powers of $\sqrt{N}$. ( $\sqrt{N}$ is the
measure of the radius of the system.) The same analytical expression
can be found just considering the sizes up to $N = 20$.

The expectation values (interaction energies per particle) of the
meron states (Eq.(\ref{mnn}) with $n = 1, 2, 3$) with respect to the
ground state value are plotted in Fig. 3 along with analytical fits.
\begin{figure}
\includegraphics[width= 7cm]{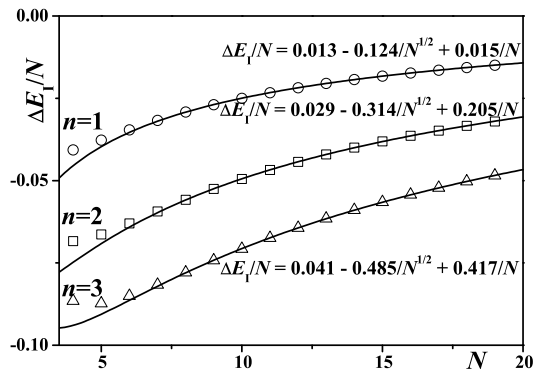}
\caption{The interaction energy per electron with respect to the
ground state value for the meron states with $n=1,2,3.$ Continuous
lines depict fits to the calculated values. Fits are
expanded in powers of $1/\sqrt{N}$ and the leading term in the
$N\rightarrow\infty$ limit is positive. This behavior leads to the
absence of merons in large systems.}
\end{figure}
Again we take $V_{1} = 1$
and there are no contributions from the first pseudopotential
$V_{0}$ because the states in Eq.(\ref{mnn}) are zero energy
eigenstates with respect to that pseudopotential. The analytical
fits are expansions in inverse powers of $\sqrt{N}$, which is the
measure of the radius of the system. The expectation values come
with overall negative sign with respect to the ground state value as
each excitation represents an inflation of the volume of the system
and therefore increase in the average distance and decrease in the
interaction among particles. Nevertheless, we can find extrapolating
the fitted analytical expressions to larger number of particles and
system sizes that the positive first term (small but always present
and approximately proportional to a constant times $n$) will
overcome all other terms in the $N \rightarrow \infty $ limit. This
leads to the conclusion that in the thermodynamic limit, $\Delta E_I
\sim N$ and positive, the merons are not present in the excitation
spectrum. Like vortices in the XY model they are confined in pairs
as skyrmion excitations.

In a strong magnetic field and realistic situation with a harmonic
confining potential, we can model  the confining part of the
Hamiltonian as
\begin{equation}
H_{C} = g \Delta M.
\label{hc}
\end{equation}
$\Delta M$ is the angular momentum measured with respect to the ground state value,
 Eq.(\ref{mom}),
and $g$ is a positive constant.
Then we may expect
(due to the positive third terms in the expressions in Fig. 3 with nearly
$n (n - 1)$ behavior)
that in the case of the Hamiltonian,
\begin{equation}
H = H_{I} + H_{C}
\label{hh}
\end{equation}
the system is prone to the instabilities (spin reconstructions)
described by the meron excitations in Eq.(\ref{mnn}). Thus we
plotted the total energy of a droplet with $N = 19$, Fig. 4,
\begin{figure}
\includegraphics[width= 7cm]{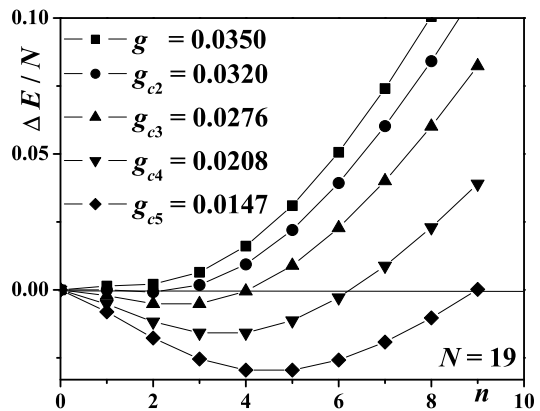}
\caption{The total energy gain per electron as a function of $n$ for
$N=19,$ and for different values of the confining constant $g$.
There is no meron ground state for $g>0.032$. Here, $g_{ci}, i=2,3,4,5$ ($c$ for critical) are calculated values of the slope of the confining potential when the total energies for creation of merons with winding numbers $n=i-1$ and $n=i$ are equal. Therefore $g_{ci}$ has the meaning of the slope of the confining potential when meron with higher winding number $i$ enters the droplet.}
\end{figure}
for a few $g_{c}$ (c - critical) when the total energy of the
excitation with fixed $n$ equals the energy of the one with $n + 1$.
This is a situation when one ground state with fixed $n$ becomes
less favorable with decreasing $g$ and substituted with another one
with the meron number equal to $n + 1$. Somehow, as an only
exception, our mean field approach predicts that the $ n = 2$ state
reconstruction precedes the one with $ n = 1 $ ( Fig. 4). This is
very likely an artifact of our choice of the interaction potential (
Eq.(\ref{iham})).  The implied ground states are only possible when
the size of the system is not large that the first and the second
term in the expansions of the interaction energies become comparable
in size. Approximately this happens when $ N \gtrsim 100$ as can be
seen from the analytical fits in Fig. 3.

With a view on the  theory of the evolution of a quantum dot with
magnetic field in Ref.~\onlinecite{saa} in terms of completely
polarized states incorporating (quasi)hole excitations, and its
reasonable agreement with experiment,~\cite{oos} the meron evolution
that we advocate would occur at much smaller gyromagnetic ratio (or
stronger interaction strength). Further merons of higher meron
number ($n$ = 2 or 3), ``giant merons'', are highly unlikely in
larger droplets, $N > 5$, as described in Ref.~\onlinecite{pm}. This will
likely cause that implied $N/2$ periodicity would be slightly
modified, as entering merons would accomodate also on orbitals away,
off the center just in the case of vortex excitations in the
completely polarized case.~\cite{saa}


To illustrate the bulk single meron configuration that the
construction in Eq.(\ref{mnn}) for $n=1$ represents we plotted in
Fig. 5 its spin density $\mathbf{S}(z),$ and charge density $\rho(z),$
as functions of the 2D coordinate $z = |z| \exp(\text{i}\varphi)$ in
 the case of $N=20.$ The analytic expressions (which are plotted) for spin density $\mathbf{S}(z)$
 are
\begin{eqnarray}
  S_z(z) &=& (N-|z|^2)f_N(|z|), \\
  S_x(z) &=& 2\sqrt{N}|z|\cos(\varphi)f_N(|z|), \\
  S_y(z) &=& 2\sqrt{N}|z|\sin(\varphi)f_N(|z|),
\end{eqnarray}
and for charge density $\rho(z)$  is
\begin{equation}
  \rho(z) = (N+|z|^2)f_N(|z|),
\end{equation}
where
\begin{equation}
f_N(|z|)=\sum_{m=0}^{N-1}\frac{|z|^{2 m}e^{-\frac{|z|^2}{2}}}{2\pi2^{m+1}m!\left(m+1+\frac{N}{2}\right)}.
\end{equation}
Therefore we have $\mathbf{S}^2(z)=\rho^2(z).$ From Fig. 5 and these expressions we can conclude that for $|z|\leq\sqrt{N}$ we have a description of a single meron that is created in the bulk of the system. Close to the boundary $\mathbf{S}(z)$ vector lies in the 2D plane and rotates for $2\pi$ as expected for a meron.

\begin{figure}
\includegraphics[width= 7cm]{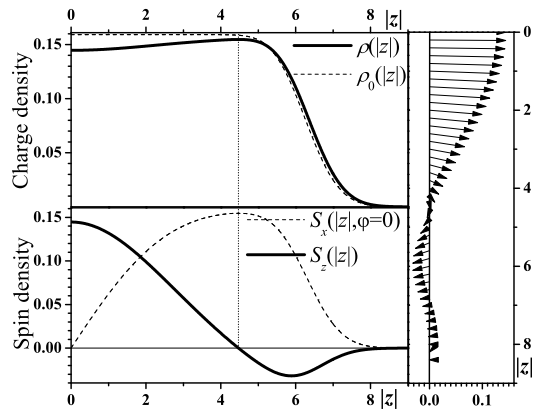}
\caption{Plotted are for $N=20$, in the upper panel, charge density $
\rho(|z|)$ with respect to the density in the
ground state,  $ \rho_{0}(|z|)$, in the lower panel, $S_{z}(|z|)$ component of the
spin density along $ S_{x}(|z|,\varphi=0) = \sqrt{S_{x}^{2}(|z|) +
S_{y}^{2}(|z|)}$, and in the side panel, spin density vector
$\mathbf{S}(|z|)$ as functions of radius $|z|$ for the meron
construction, Eq.(\ref{mnn}) with $n = 1$. A small dip in the charge
density for $|z| < \sqrt{N}$ describes the deficit of the charge due to the presence of a meron.
The amount of the missing charge is 0.486e, close to e/2 as expected for a meron.}
\end{figure}

Now, we must pose a question which quantum mechanical states - eigenstates of
$M$ and $S_{z}$, correspond to the "mean-field" or "classical" constructions expressed
by Eq.(\ref{mnn}).
We will take that each meron classical construction  corresponds to the
quantum mechanical state that we find by expanding the spinor part of Eq.(\ref{mnn})
with eigenvalue of $M$ corresponding to the expectation value in Eq.(\ref{mom}).
On the other hand, the states that were proposed for  spin (global not only edge) reconstructions
in Ref.~\onlinecite{oak} are of the following form
$(\Sigma_{1}^{\dagger})^{k_{o}} |C_{N}>$.
$|C_{N}>$ denotes the filled with $N$ spin $\uparrow$
 particles LLL - a Vandermonde determinant, and
$ \Sigma_{1}^{\dagger} = \sum_{m} \sqrt{m + 1} \; c_{m + 1, \downarrow}^{\dagger} c_{m, \uparrow}$,
an exciton operator with $c_{m}^{\dagger}$ and $c_{m}$, the electron
creation and annihilation operators, and $m$ denoting
single-particle angular momentum.  $ k_{o} = N/2$, if $N$
is even,
and after a little inspection we can find out that the state described by the formula
coincides with
our quantum mechanical eigenstate analog of
the construction in Eq.(\ref{mnn})
 with $n$ equal to one.
As we now show explicitly, in the case of the merons of
Eq.(\ref{mnn}), these constructions carry very small $S_{z}$. The
states in Eq.(\ref{mnn}) are eigenstates of $M + n S_{z} - M_{o}$
with eigenvalues $n N/2$.~\cite{mvm} Therefore $\langle S_{z}\rangle $ in
these states is $\langle S_{z}\rangle = - (n - 1)/2$ and $n\geq1 $ for
merons.

To find out whether even in small droplets the mean field ansatz has
a distribution of participating eigenstates of $M$ (and $S_{z}$)
very much peaked around the quantum mechanical states with $\Delta M$ given in Eq.(\ref{mom}),
we calculated these distributions (Fig. 6) in the cases of a droplet
with $N = 17$ and $N = 20$. Although our trial functions are not eigenstates of
the angular momentum, the calculated participation ratios,
\begin{eqnarray}
\label{pr}
\frac{P_{k}}{P_{\text{max}}}&&=\left(\frac{2}{N}\right)^{k n}\times\nonumber\\
&&\sum_{<i_0,\dots,i_{N-1}>}(0+i_0n)!\cdots(N-1+i_{N-1}n)!,\phantom{mm}
\end{eqnarray}
where $i_f=0,1$ and $\sum_{f=0}^{N-1}i_f=k$, are indeed well peaked
around expectation values,
\begin{equation}
\label{exp}
\Delta M(N, n)=\frac{\sum_{k=0}^Nn k P_{k}}{\sum_{k=0}^NP_{k}}.
\end{equation}
\begin{figure}
\includegraphics[width= 7cm]{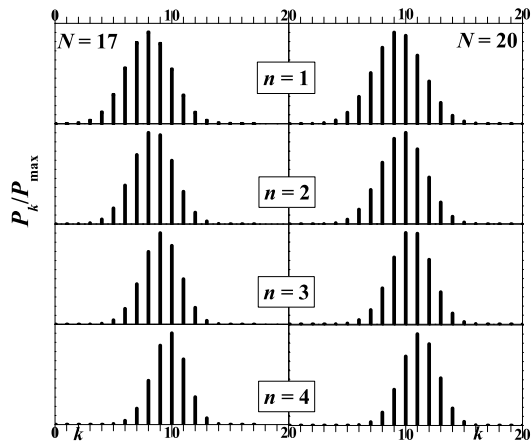}
\caption{Participation ratios $P_k/P_{\text{max}}$ of the
eigenstates of the angular momentum with eigenvalues $M=N(N-1)/2+nk$
as functions of $k=1,\dots,N$ for the meron construction defined by Eq.(\ref{mnn}).}
\end{figure}

\section{Discussion and Conclusions}
The meron physics we described above should be in
principle detectable in lateral quantum dots, in which interaction
effects are strong, in their not yet explored regime beyond the MDD
state,~\cite{ni} as well just below the MDD state where depolarized
$(S = 0)$ states were already discovered.~\cite{haw}

Besides the study in Ref.~\onlinecite{pm} of small systems, there are
studies in Refs.~\onlinecite{sil,pee1,pee2} without the LLL approximation
that find depolarized states that we can identify as meron ground
states. They appear at angular momenta, multiplets of $N/2$,
for $N$ even.


Here we addressed the question of the existence of the meron ground
states in larger quantum Hall droplets. Our conclusion is, as long
as the Zeeman term is small, the meron ground states are viable
solutions of these droplets for $N $ smaller than $\sim 100$. Also
we expect the appearance of merons in any small enough system of
particles with the lowest Landau level quantization and a
degenerate, additional degree of freedom. That may happen in
graphene structures in the presence of magnetic field (with valley
degree of freedom) or rapidly rotating quantum gases (with spin
degree of freedom).

A previous study, Ref.~\onlinecite{yzdw}, of few-electron qnatum dots,
without the Zeeman term, was based on restricted (conditional) wave
function (RWF) method.~\cite{sat} As the method underlines explicit
fixing of the spin projection of each electron, the vortices
observed are analogous to those of polarized systems. A more
detailed and recent study in Ref.~\onlinecite{sarei} is based on the same
method and discusses the problem of two distinguishable species of
bosons or fermions which rapidly rotate. As usual in these
circumstances the LLL approximation is applied and the fermion
system is (up to the difference caused by the distinguishability)
the same as in the case of the quantum Hall droplet discussed in
this paper. The first vortices that enter the MDD come at the
increase of angular momentum of $N/2$. Due to being just
vortices of one kind of fermions, a density of fermions of another
kind at their core was detected. Because of this phenomenon these
vortices are called coreless vortices. For a larger droplet $(N =
20)$ it was demonstrated using DFT (density functional theory) and
the RWF method that a coreless vortex enters the droplet (at an
angular momentum slightly higher than that of MDD). Based on these
observations, i.e. that coreless vortices as merons show an increase
of polarization at their core (center), and our studies we can
conclude that merons are the solutions of the quantum Hall droplet
in the fully spin rotationally invariant case, i.e. in an
indistinguishable picture, and correspond to these coreless vortices
in the distinguishable picture.
\section{ACKNOWLEDGMENTS}

This work was supported by Grants No. 141035, 141017, and 141014 of the
Serbian Ministry of Science.
\appendix*
\section{}
First we consider a fixed spin configuration - component of the total
wave function which expectation value of the interaction energy we
want to calculate:
\begin{equation}
\Psi(z_{1},\ldots,z_{N}).
\end{equation}
$\Psi$ in general does not have any overall definite symmetry property (i.e.
it is not antisymmetric or symmetric in general). We first have to perform
the calculation for each component $\Psi$ separately and then add contributions.
We do the calculation by performing the projection to angular momentum,
$ l = 1$, component of wave function $\Psi$ for each pair of particles
$z_{i}$ and $z_{j}$. The component can be extracted by expanding $\Psi$
in powers of $(z_{i} + z_{j})$ and $(z_{i} - z_{j})$ i.e.
\begin{eqnarray}
\Psi = && \sum_{m,n = 0} (z_{i} - z_{j})^{m}(z_{i} + z_{j})^{n} \times \nonumber\\
&&C_{m,n}(z_{1},\ldots,z_{i-1},z_{i+1},\ldots,z_{j-1},z_{j+1},\ldots,z_{N}),\nonumber\\
&&
\end{eqnarray}
and taking only $m = 1$ contribution - wave function in the expansion.
Because we take $V_{l=1} = 1$, the normalized, integrated over all
coordinates square of the projection would be the contribution to the
interaction energy from the given spin configuration $\Psi$ and fixed
pair of particles.

\end{document}